\documentclass[11pt,a4paper]{article}

\usepackage{amssymb}
\usepackage{amsmath}

\newcommand{\pa}{\partial}
\newcommand{\Rset}{{\mathbb R}}

\newcommand{\Nset}{{\mathbb N}}

\newcommand{\myref}[1]{(\ref{#1})}

\newcommand{\om}{\omega} 

\newcommand{\de}{\delta}
\newcommand{\al}{\alpha}
\newcommand{\eps}{\epsilon}
\newcommand{\la}{\lambda}
\newcommand{\ga}{\gamma}

\renewcommand{\leq}{\leqslant}
\renewcommand{\geq}{\geqslant}

\newcommand{\sur}[2]{{\displaystyle\mathop{#1}_{#2}}}

\newlength{\somme}
\newlength{\sommep}
\newcommand{\demi}{\frac{1}{2}}

\usepackage{amssymb}
\usepackage{amsmath}
\usepackage{epsfig}
\usepackage{latexsym}
\usepackage{color}

\newcommand{\tla}{\widetilde{\la}}
\newcommand{\teps}{\tilde{\varepsilon}}
\newcommand{\tg}{\widetilde{\ga}}
\newcommand{\ft}{\widehat{\pi}}

\renewcommand{\eps}{\varepsilon}
\renewcommand{\geq}{\geqslant}
\renewcommand{\leq}{\leqslant}

\newcommand{\pilav}{\ft_{v_0}(\la)}
\newcommand{\piboth}{\ft_{(v_0)}(\la)}
\newcommand{\rz}{\mathbf{r_0}}
\newcommand{\ru}{\mathbf{r_1}}

\begin{document}
\title{Injected Power Fluctuations in Langevin Equation}
\author{Jean Farago\footnote{e-mail: farago@lps.ens.fr}\\
\textit{Laboratoire de Physique Statistique, \'Ecole Normale
Sup\'erieure\footnote{CNRS, UMR 8550}}\\
\textit{24 rue Lhomond, 75231 Paris cedex 05, France.}}
\date{}
\maketitle
\abstract{In this paper, we consider the Langevin equation from an unusual  point of view, that is as an archetype for a dissipative system driven out of equilibrium by an external excitation. Using path integral method, we compute exactly the
probability density function of the power (averaged over a  time interval of length
$\tau$) injected (and dissipated) by the random force 
into a Brownian particle driven by a Langevin equation. The resulting distribution, as well as the associated large
deviation function, display strong asymmetry, whose origin is
explained. Connections with  the so-called ``Fluctuation Theorem'' are thereafter discussed. Finally, considering Langevin equations with a pinning potential, we show that the large deviation function  associated with the injected power is \textit{completely} \textit{insensitive} to the presence of a potential.}

{\bf Keywords:} Fluctuation phenomena, random processes, noise and
Brownian motion.

\section{Introduction}

Usually, works concerning out of equilibrium stationary systems deal
with their local statistical properties, some assumptions of
homogeneity and isotropy being reasonably assumed or understood:
traditional theory of turbulence, which focuses on local correlations
of the velocity field furnishes a good illustration of that. A rather new
way to study such systems was recently proposed by some authors \cite{labbe,aumaitreRB,aumaitreVK}, who
preferred concentrate their efforts to characterize the process of
injection of energy, imperatively required to sustain the stationary
state. To be more precise, in numbers of such situations, there exists
a channel of energy injection, usually located at boundaries of the
system (rotating blades driving a turbulent flow, heated plate in
Rayleigh-B\'enard convection, piston shaking granular matter at an
edge of a vessel, etc\ldots) together with a distinct channel of
dissipation, often provided by a bulk dissipation mechanism, viscosity
or inelastic collisions. This duality is explicitly expressed in the
dynamical equation for the energy which can always be written in the
form $\dot{E}=\mathcal{I}-\mathcal{D}$, where $\mathcal{D}$ is proportional to a coefficient of
dissipation whereas $\mathcal{I}$ is entirely due to the injection
process: for instance, the evolution of the kinetic  energy of an
uncompressible fluid is given by
\begin{align}
\pa_t\left[\rho\int_V\mathbf{v}^2dV\right]=\int_{\pa V}(\eta
v_i\sigma_{ij}n_j-(p+\rho \mathbf{v}^2/2)\mathbf{v.n})dS-\frac{\eta}{2}\int_V\sigma_{ij}\sigma_{ij}dV,
\end{align}
where one recognizes easily a surface injection term and a dissipative
bulk term ($\sigma_{ij}=\pa_iv_j+\pa_jv_i$)
. These two distinct ``gates'' lead to the establishment of a
permanent flow of energy throughout the system, and is obviously a
primordial feature  amongst the stationary properties of the
system. Thus, some experimental measurements
\cite{labbe,aumaitreRB,aumaitreVK} and numerical simulations
\cite{aumaitre} were performed to characterize the injection of energy
(more easily reachable than the dissipation), or more precisely the
probability density function (pdf) $\pi(\eps)$ of
$\eps=\frac{1}{\tau}\int_0^\tau \mathcal{I}(t)dt$, the averaged
injected power during a time interval of length $\tau$. In particular, in some
works \cite{aumaitre,ciliberto,sasa}, the quantity
\begin{align}
\rho(\eps)=\frac{1}{\tau}\log\frac{\pi(\eps)}{\pi(-\eps)}
\end{align}
was measured, and it was noticed that, according to the
\textit{conclusions} of the so-called ``Fluctuation Theorem''
\cite{ecm,gc}, $\rho(\eps)$ seemed to tend to a straight line for
large $\tau$. These observations were quite surprising, for the
Fluctuation Theorem was established for time-reversal systems, a
property which is at the heart of the demonstration of the
theorem. A convincing argument was recently proposed \cite{aumaitre}
to explain this seemingly universal behaviour: if one considers that
the signal of $\mathcal{I}(t)$ has only finite correlations, large deviation theory
predicts that $\log \pi(\eps)\sim \tau f(\eps)$ for large
$\tau$. Consequently, $\rho(\eps)=f(\eps)-f(-\eps)$; but for large
$\tau$, it is extremely unprobable to observe  large \textit{negative}
occurences of the averaged injected power (but not unpossible), since
in average, the injected power in a dissipative system is positive. As
a result, in concrete measurements of $\rho(\eps)$, one can presumably
only measure $\rho$ in a short vicinity of zero, for which
$f(\eps)-f(-\eps)\approx2f'(0)\eps$ with an excellent approximation:
that is why a straight line behaviour is always measured in real or
numerical experiments.

Nevertheless, there was a lack of a real example of a nonequilibrium
dissipative model in which $\pi(\eps)$ could be fully computed, in order to
show that time-reversal symmetry is absolutely required for the
Fluctuation Theorem ``really'' to hold.
The initial aim of this paper was thus to provide a simple model, where the
pdf $\pi(\eps)$ is exactly computable, and which mimics as simply as
possible the presence of two distinct channels for energy flow. One of
the simplest (nontrivial) systems fulfilling these requirements is provided by the Langevin equation
\begin{align}\label{langevinlibre}
  \dot{v}+\ga v&=\psi(t)\\
<\psi(t)\psi(t')>&=2D\de(t-t')
\end{align}
\textit{where fluctuations $\psi(t)$ and dissipation $\ga v$ are considered as two different sources of modification of energy} :
\begin{align}
\frac{d}{dt}\left(\demi v^2\right)&=\underbrace{-\ga v^2}_{\mbox{dissipation}}+\underbrace{\psi v}_{\mbox{injection}}
\end{align}
Note that in absence of $\psi$, the system is clearly dissipative and
non time-reversal invariant, a property also shared by realistic
hydrodynamic or granular systems.

This interpretation of the Langevin equation is clearly uncommon: 
 deriving  the Langevin equation as an evolution equation for a
Brownian particle in a thermalized surroundings, it appears clearly
that the dissipative term $-\ga v$ and the fluctuating term $\psi$ are
two different faces of the action of the reservoir on the particle;
the fluctuation-dissipation relation $D=\ga k_BT$ testifies this
profund link. In the present case however, we consider the Langevin
equation as a given evolution equation, \textit{irrespective of its physical
origin}, and interpret it as a dissipative system $\dot{v}+\ga v=0$
shaked by a random gaussian force $\psi$; in particular, we will make
in the following no reference  to the fluctuation-dissipation relation just cited.

\medskip

Thus, we compute in this paper  the pdf of 
\begin{align}
\eps=\frac{1}{\tau}\int_0^{\tau}dt'\psi(t')v(t')
\end{align}
in the permanent regime using the path integral method (we compute also  the pdf
of the dissipated power). 
We show as expected that the function $\rho(\eps)$ is in this case not a straight line (to prevent any confusion, it is worth noting that our result is not contradictory with that of Kurchan \cite{kurchan}, who proved the Fluctuation Theorem for the power injected by an \textit{external} operator acting on a Brownian particle: the physical situation considered here is by no means the same), though this nonlinear behaviour  is extremely difficult to verify numerically.

But some other interesting and novel features emerge also from our study, which concern merely the large deviation function $f(\eps)$ associated with $\pi(\eps)$: we show that this function is not a regular function but displays a unexpected  second order singularity; we discuss its physical origin and show that it is intimately associated with the permanent regime which allows rare initial fluctuations of the velocity which have deep consequences for the large deviation function.

Finally, we study also the effect of a pinning potential on this large deviation function and show that adding a potential does not have rigorously any consequence on the large deviation function. This ``universality'' confirms the relevance of
 considering the averaged
injected power as a probe for extracting global features of energy flow into a
nonequilibrium system.

\section{Free Brownian motion}

\subsection{Characteristic functions}

We consider a particle of mass $1$, velocity $v$, whose dynamics is given by \myref{langevinlibre}, and want to compute the pdf of $\eps$ in the permanent regime. To do that it is convenient to compute first its characteristic function 
\begin{align}
\ft(\la)=\left\langle e^{-\la\tau\eps}\right\rangle
\end{align}
which is related to the Fourier transform of $\pi(\eps)$ by
$FT[\pi(\eps)](k)=\ft(-ik/\tau)$. In some works
\cite{derrida}, one computes already at this stage the
asymptotic exponential dependence of the characteristic function:
$\langle e^{-\la\tau\eps}\rangle\sim e^{\tau g(\la)}$, and retrieves
$f(\eps)$ as the inverse Legendre transform of $g(\la)$:
$f(\eps)=g(\la)+\la \eps,\ \ g'(\la)=-\eps$. We will see that this procedure is not appropriate
here, for reasons which will be made clear later. One prefers thus compute first exactly $\ft(\la)$, what is here fortunately feasible.

Let us consider the stochastic equation \myref{langevinlibre}. The propagator $P(v_1,\tau|v_0,0)$ can be easily expressed in terms of a path integral \cite{wiegel, feynman,zinnjustin}:
\begin{align}
P(v_1,\tau|v_0,0)=e^{\frac{\ga}{2}\tau}\times\int_{v(0)=v_0}^{v(\tau)=v_1}[\mathcal{D}v] \exp\left(-\frac{1}{4D}\int_0^{\tau}dt\left(\dot{v}+\ga v\right)^2\right)
\end{align}
From this formula, one can deduce the probability density associated with a given path:
\begin{align}
\mathcal{P}([v(u), 0\leq u\leq \tau])= \exp\left(\frac{\ga\tau}{2}-\frac{1}{4D}\int_0^{\tau}dt\left(\dot{v}+\ga v\right)^2\right)
\end{align}
As a result, one has
\begin{align}
\pilav&\equiv\left\langle e^{-\la\tau\eps}\right\rangle_{v_0}=\int_{-\infty}^{\infty} dv_1 \int_{v(0)=v_0}^{v(\tau)=v_1} [\mathcal{D}v] \mathcal{P}([v]) \exp\left(-\la\int_0^\tau v(\dot{v}+\ga v)\right)\\
&=e^{\frac{\ga\tau}{2}}\int_{-\infty}^{\infty}\!\!\!\!\!dv_1e^{-\left(\frac{\la}{2}+\frac{\ga}{4D}\right)(v_1^2-v_0^2)}\int_{v(0)=v_0}^{v(\tau)=v_1}\!\!\!\!\!\!\![\mathcal{D}v]\exp\left(-\frac{1}{4D}\int_0^\tau [\dot{v}^2+(\ga^2+4D\la\ga)v^2]\right)\label{interm1}
\end{align}
where $\langle\ldots\rangle_{v_0}$ designates an average over the realizations of $v$ such that $v(0)=v_0$. The path integral in \myref{interm1} is well-known and its value can be exactly computed \cite{wiegel}:
\begin{multline}
\int_{v(0)=v_0}^{v(\tau)=v_1}\!\!\!\!\!\!\![\mathcal{D}v]\exp\left(-\frac{1}{4D}\int_0^\tau [\dot{v}^2+\al^2 v^2]\right)=\\
\left(\pi\frac{4D}{\al}\sinh\al\tau\right)^{-1/2}\exp\left(-\frac{\al}{4D}\frac{(v_1^2+v_0^2)\cosh(\al\tau)-2v_0v_1}{\sinh\al\tau}\right)
\end{multline}
Thus, let us define
\begin{align}
\tg&=\ga\tau\\
 \tla&= 2D\la/\ga\\
 \eta&=\sqrt{1+2\tla}
\end{align}
one has
\begin{align}
\pilav&=e^{\tg/2}\left(\cosh\eta\tg+\frac{1+\tla}{\eta}\sinh\eta\tg\right)^{-1/2}\times\exp\left(\frac{v_0^2\ga}{2D}\frac{\tla^2/2}{\eta\coth\eta\tg+1+\tla}\right)\label{pilav0}
\end{align}
To get $\ft(\la)$, one has to average over $v_0$ with a probability $\propto \exp(-v_0^2\ga/2D)$, since it is the distribution of $v(t)$ in the permanent regime:
\begin{align}
\ft(\la)&=\sqrt{\frac{\ga}{2D\pi}}\int dv_0\ e^{-v_0^2\ga/2D}\left\langle e^{-\la\tau\eps}\right\rangle_{v_0}\\
&=e^{\tg/2}\left(\cosh\eta\tg+\frac{1+\tla-\tla^2/2}{\eta}\sinh\eta\tg\right)^{-1/2}\label{pila}
\end{align}
The calculation leading to \myref{pilav0} and \myref{pila} assumes \textit{a priori} $\la>0$, but it is useful to study the properties of the analytical continuations of these formulae. First, one remarks that the square root in the definition of $\eta$ does not induce any breaking of analyticity, for it appears always in quantities such $\cosh (\eta\tg)$ or $\sinh(\eta\tg)/\eta$ which are entire functions of $\la$. Thus, one will assume in the following that $\eta=x+iy$ has \textit{positive} real part.

Let us first look at $\pilav$ for $v_0=0$ (the exponential term will not modify the analytical properties of $\pilav$). It is astute to write it as\footnote{this manipulation makes the leading term in the limit of large $\tau$ explicit and allows for a simple localization of the cuts.}
\begin{align}
\left\langle e^{-\la\tau\eps}\right\rangle_{0}&=e^{(1-\eta)\tg/2}\left(\frac{1+e^{-2\eta\tg}}{2}\right)^{-1/2}\times\left(1+\demi\left(\eta+\frac{1}{\eta}\right)\tanh\eta\tg\right)^{-1/2}\label{cidessus}
\end{align}
and assume again a cut in the negative real semi-axis for the
definition of the square root. Breaks in analyticity can arise if only
one  of the two terms of \myref{cidessus} (separated by ``$\times$'')
is not defined (a superposition of two cuts restores the analyticity,
since they are associated with a square root). The last term is not defined for $\eta=x+iy$ such that 
\begin{align}
x=0\ \ \text{and}\ \ 1<\demi\left(y-\frac{1}{y}\right)\tan\tg y
\end{align}
This induces in the $\tla$-space a dashed half cut localized in the negative real axis, beginning at the value $\tla_-$ of $\tla$ less but closest to $-\demi$ such that
\begin{align}
1=\frac{1}{2}\left(\sqrt{|1+2\tla_-|}-\frac{1}{\sqrt{|1+2\tla_-|}}\right)\tan\tg\sqrt{|1+2\tla_-|}
\end{align}
As $\tg\propto\tau$,   it is clear that
$\tla_-\rightarrow -\demi$ when $\tau\rightarrow\infty$.

The first term $e^{(1-\eta)\tg/2}\sqrt{2/(1+e^{-2\eta\tg})}$ ,  as a
function of $\eta$, is analytical in the region $x\geq 0$ (except at
points $x=0,y\equiv\frac{\pi}{2}[\pi]$). But if one considers it now
as a function of $\tla$, it appears cuts in the $\tla$-space due to the prescription $x=\text{Re}(\eta)>0$. It is easy to verify that these cuts are defined by
\begin{align}
(1+2\tla)\in\frac{1}{\tg}[-(3\pi/2+2k\pi)^2,-(\pi/2+2k\pi)^2],\ \ \text{for $k\in\Nset$}
\end{align}
and again are located in the $\tla<-\demi$ half axis.

To summarize, $\pilav$ continued on the whole complex plane (as a
function of $\la$) has a cut, dashed line shaped\footnote{We do not
give further details on the precise structure of this ``hacked'' cut,
for they are of no importance in  the following.} and localized  on
the negative real axis. It begins at a value $\tla_-$ less than $-\demi$, but  tends to $-\demi$ for large $\tau$.

For $\ft(\la)$, the situation is similar, and gives also this dashed negative cut. But there is a fundamental discrepancy, for the ``second term''  $[1+(1+\tla-\tla^2/2)\tanh(\eta\tg)/2\eta]^{-1/2}$ gives here a novel cut, plain and localized on the \textit{positive} real axis, and beginning at a value $\tla_+$ solution of
\begin{align}
\left(\left[\sqrt{1+2\tla_+}\right]^{3}-6\sqrt{1+2\tla_+}-\frac{3}{\sqrt{1+2\tla_+}}\right)\times\tanh\tg\sqrt{1+2\tla_+}=8
\end{align}
When $\tau\rightarrow\infty$, one has simply $\tla_+\rightarrow 4$.
One has summarized these analytical properties on figure \ref{anaprop}; it is worth noticing that the extra cut of $\ft(\la)$ has deep consequences on the shape of the large deviation function, as we show in the following.
\begin{figure}
\begin{center}
\input{cutinlambda.pstex_t}
\end{center}
\caption{Cuts of functions (a): $\pilav$  and (b): $\ft(\la)$  in the complex $\tla$ plane (see text for details).}\label{anaprop}
\end{figure}

\subsection{Pdf of injected power and large deviation functions}

Why did we study analytical properties of $\pilav$ and $\ft(\la)$ ? A
priori, we could have remarked that $\pilav$, as well as $\ft(\la)$
(we will use henceforth the notation $\piboth$ to designate both
$\pilav$ and $\ft(\la)$) are such that
\begin{align}
\piboth &\sur{\sim}{\tau\rightarrow\infty} \exp\left[\tau
g(\la)\right]\\
\text{with}\ \ \  g(\la)&=\frac{\ga}{2}\left(1-\sqrt{1+2\tla}\right)
\end{align}
(where ``$\sim$'' means an equivalence between logarithms
\cite{derrida}), and using a traditional recipe, obtain the large
deviation function of $\pi_{(v_0)}(\eps)$ as the inverse Legendre transform of
$g(\la)$. This is actually correct for $\pi_{v_0}(\eps)$, but
gives wrong results for $\pi(\eps)$, because the above mentioned
procedure neglects completely the presence of the extra cut in
$\pi(\eps)$, whose origin is the prefactor of the exponential leading
term.

Thus, it is more suited to first express the Fourier inversion of
$\piboth$ properly, and only thereafter extract the associated large
deviation function from a saddle point expansion \cite{benderorszag}.

Let us define $\teps=\eps/D$ the dimensionless injected power. One has
\begin{align}
\pi_{(v_0)}(\teps)&\equiv D\times\pi_{(v_0)}(\eps)\\
&=\frac{\tg}{4i\pi}\int_{-i\infty}^{i\infty}d\tla\ \ \ft_{(v_0)}(\ga\tla/2D)\exp\left(\frac{\tg}{2}\teps\tla\right)
\end{align}
 The leading exponential term in this integral
is $\exp[\tau h(\tla)]$ with
$h(\tla)=\frac{\ga}{2}(\teps\tla+1-\sqrt{1+2\tla})$. The saddle point
expansion method requires to distort the usual path of integration
$\tla\in i\Rset$ in such a way that $h(\tla)$ be always real. Let us
parametrize $\sqrt{1+2\tla}=x+iy$ (with $x\geq 0$). Two paths ensure
$\text{Im} h(\tla)=0$; they are
\begin{align}
\tla_1(x)&=\frac{x^2-1}{2}\ ,\ \ \ (x\in[0, \infty[)\ \ \ \text{and}\\
\tla_2(y)&=\frac{1}{2}\left(\frac{1}{\teps^2}-y^2-1\right)+i\frac{y}{\teps}\
,\
\
\ (y\in]-\infty,+\infty[)
\end{align}
\begin{figure}
\begin{center}
\input{saddlepaths.pstex_t}
\end{center}
\caption{Paths $\tla_1$ and $\tla_2$ where equation $\mbox{Im}[g(\tla)]=0$ is
verified for $\teps>0$. One indicates also the schematic behaviour of
$\mbox{Re}[g]$ along these paths : if it grows to $\infty$ (``up'') or
decreases to $-\infty$ (``down''). The saddle is located at the crossing of the
two paths. Note that $\tla_2$ does no longer exist 
if $\teps\leq0$, but in that case $\tla_1$ vanishes at infinity.}\label{saddlepaths}
\end{figure}
(see figure \ref{saddlepaths}) and give $h(\tla_1(x))=\frac{\ga}{2}[\frac{\teps}{2}(x^2-1)+1-x]$ and
$h(\tla_2(y))=-\frac{\ga}{2}[\frac{\teps}{2}y^2+\frac{1}{2\teps}(\teps-1)^2]$
respectively.

Owing to the prescription $\text{Re}\sqrt{1+2\tla}\geq 0$, the path
$\tla_2$ does exist only if $\teps>0$. The next step consists in choosing the
right path of integration. From now, the computations for
$\pi_{v_0}(\eps)$ and $\pi(\eps)$ differ.

\bigskip

For $\pi_{v_0}(\teps)$, the situation is relatively simple, since neither
$\tla_1$ nor $\tla_2$ crosses the cut-off of $\ft_{v_0}$. In the
following, we give details only for $v_0=0$ because it simplifies a
bit the computation; we postpone remarks concerning the incidence of a
non-zero initial velocity. 

It is
easy to see that $\tla_2$ is a valid integration path if
$\teps>0$ (in particular, the prefactors of the exponential do not
cause problems of convergence). If  $\teps\leq 0$, the path $\tla_2$
is not defined, but using a semicircular contour, one shows easily
that (evidently) $\pi_0(\teps)=0$ strictly in this case. To summarize, and after
some calculations, the pdf $\pi_0(\teps)$ of the dimensionless injected power
$\teps=\frac{1}{D\tau}\int_0^\tau dt \psi(t)v(t)$, \textit{knowing that
the initial velocity $v_0$ is zero}, is given by
\begin{align}\label{pizero}
\pi_0(\teps)&=\left\{\begin{array}{ll}
I(\teps)\times e^{-\frac{\tg}{4\teps}(\teps-1)^2}& \text{if $\teps>0$}\\
0 & \text{if $\teps\leq0$}
\end{array}\right.
\end{align}
with
\begin{align}
I(\teps)&\equiv
\frac{\tg}{4i\pi}\times\int_{\sigma-i\infty}^{\sigma+i\infty}d\eta\
\eta\ e^{\frac{\tg\teps}{4}(\eta-\teps^{-1})^2}\left(\frac{1+e^{-2\eta\tg}}{2}\right)^{-1/2}\times\left(1+\demi\left(\eta+\frac{1}{\eta}\right)\tanh\eta\tg\right)^{-1/2}
\end{align}
($\sigma$ is any positive number). It is quite difficult to simplify
this prefactor. In the large $\tau$ limit, it is equivalent to
\begin{align}
I(\teps)\sur{\sim}{\tau\rightarrow\infty}\sqrt{\frac{\tg}{\pi}}\frac{1}{\teps(\teps+1)},
\end{align}
but this equivalence is not uniformly valid for all $\teps$. In
particular, for fixed $\tau$, the large $\teps$ values give a
$\teps^{-3/2}$ regression instead (a regime which arises for
$\teps\gtrsim\tg$). 

From \myref{pizero} one gets the following large deviation function
\begin{align}
f_{0}(\teps)&=-\frac{\ga}{4\teps}(\teps-1)^2\times\theta(\teps)
\end{align}
($\theta$ is the Heaviside function), what matches the rapid evaluation above mentioned.

How are modified these results if $v_0\neq 0$ ? Essentially, an
``energetic'' initial condition gives rise to a small interval of
possible negative power injection. To be precise, if $v_0\neq 0$, the
probability $\pi_{v_0}(\eps)$ is no longer zero if
$\eps\in]-|v_0|/2\tau,0[$ (of course the positive part of the pdf is
also slightly modified). But these modifications are minor, and in
particular, they are \textit{unable} to affect the associated large deviation
function (the negative window vanishes when $\tau\rightarrow\infty$).

\bigskip

\begin{figure}
\begin{center}
\input{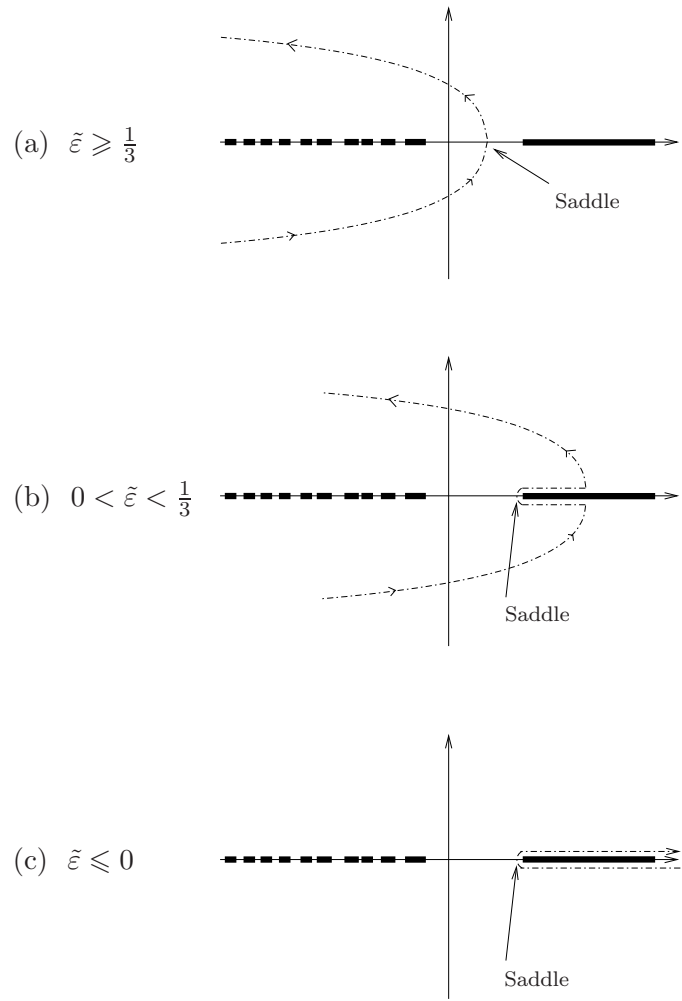}
\end{center}
\caption{Sketch of different real paths, according to the values of
$\teps$. The positive real cut fixes the location of the saddle at his
extremity $\tla_+$  as soon as $\teps<1/3$.}
\label{chemins}
\end{figure}

Let us now look at $\pi(\eps)$. As the function $h$ is the same for
both $\pi_{v_0}(\eps)$ and $\pi(\eps)$, the paths $\tla_1$ and
$\tla_2$ are similarly defined. The essential difference comes from the
positive cut in the analyticity of $\ft(\la)$, which can cross the
steepest descent paths $\tla_1$ and $\tla_2$ (see figure
\ref{chemins}).
If $\teps>1/3$, the path $\tla_2$ avoids the cuts and can be chosen as
a valid integration path (fig. \ref{chemins} (a)). On the contrary, if
$0<\teps<1/3$, it crosses the positive real cut and a portion of
the cut must be crawled along to close the path (see fig. \ref{chemins}
(b)). If $\eps<0$, the parabola $\tla_2$ does no longer exist, but
$\tla_1$ is valid (more precisely a U-shaped path sticked  on each
side of the cut -- see fig. \ref{chemins} (c)) and leads to a non zero
result for the probability. After some computations, one can deduce
the following result:
\begin{align}
\pi(\eps)&=\left\{\begin{array}{ll}
J(\teps)\times e^{-\frac{\tg}{4\teps}(\teps-1)^2} & \text{if $\teps>1/3$}\\
J(\teps)\times e^{-\frac{\tg}{4\teps}(\teps-1)^2}+K(\teps)\times e^{\tg(2\teps-1)} & \text{if $0<\teps\leq 1/3$}\\
K(\teps)\times e^{\tg(2\teps-1)} & \text{if $\teps\leq 0$}\end{array}\right.
\label{fundamental}
\end{align}
with
\begin{align}
J(\teps)&\equiv
\frac{\tg}{4i\pi}\int_{\sigma-i\infty}^{\sigma+i\infty}\!\!\!\!\!\!\!\!\!\!d\eta\
\eta\
e^{\frac{\tg\teps}{4}(\eta-\teps^{-1})^2}\left(\frac{1+e^{-2\eta\tg}}{2}\right)^{-\frac{1}{2}}\hspace*{-0.3cm}\times\left(1-\frac{1}{8\eta}\left(\eta^4-6\eta^2-3\right)\tanh\eta\tg\right)^{-\frac{1}{2}}\label{j1}\\
K(\teps)&\equiv
e^{-\frac{3\tg}{4}(3\teps-2)}\frac{\tg}{2\pi}\int_{x_+}^{1/\text{Max}(0,\teps)}\!\!\!\
\hspace{-1cm}dx\
x\ e^{\frac{\tg}{2}[\frac{\teps}{2}x^2-x]}\left(\frac{1+e^{-2x\tg}}{2}\right)^{-\frac{1}{2}}\hspace*{-0.3cm}\times\left(\frac{1}{8x}\left(x^4-6x^2-3\right)\tanh
x\tg-1\right)^{-\frac{1}{2}}\label{k1}
\end{align}
where $x_+$ is positive, defined by $\la_+=(x_+^2-1)/2$ (note
$x_+\rightarrow3$ for large $\tau$). For large values of $\tau$, one
can simplify a bit these formul\ae:
\begin{align}
J(\teps)&\sur{\sim}{\tau\rightarrow\infty} \frac{\tg}{i\pi}\int_{\sigma-i\infty}^{\sigma+i\infty}\!\!\!\!\!\!\!\!\!\!d\eta\
\eta\
e^{\frac{\tg\teps}{4}(\eta-\teps^{-1})^2}\sqrt{\frac{\eta}{(3-\eta)(\eta+1)^3}}\label{j2}\\
K(\teps)&\sur{\sim}{\tau\rightarrow\infty}\frac{2\tg}{\pi}\int_{0}^{1/\text{Max}(0,\teps)-3}\frac{dx}{\sqrt{x}}\
\left(\frac{x+3}{x+4}\right)^{\frac{3}{2}}\
e^{\frac{\tg}{2}[\frac{\teps}{2}x^2-x(1-3\teps)]}\label{k2}
\end{align}
but, as for $I$, the latter simplification for $J$ is not uniformly valid for all
concerned values of $\teps$.

\begin{figure}
\begin{center}
\resizebox{10cm}{!}{\includegraphics{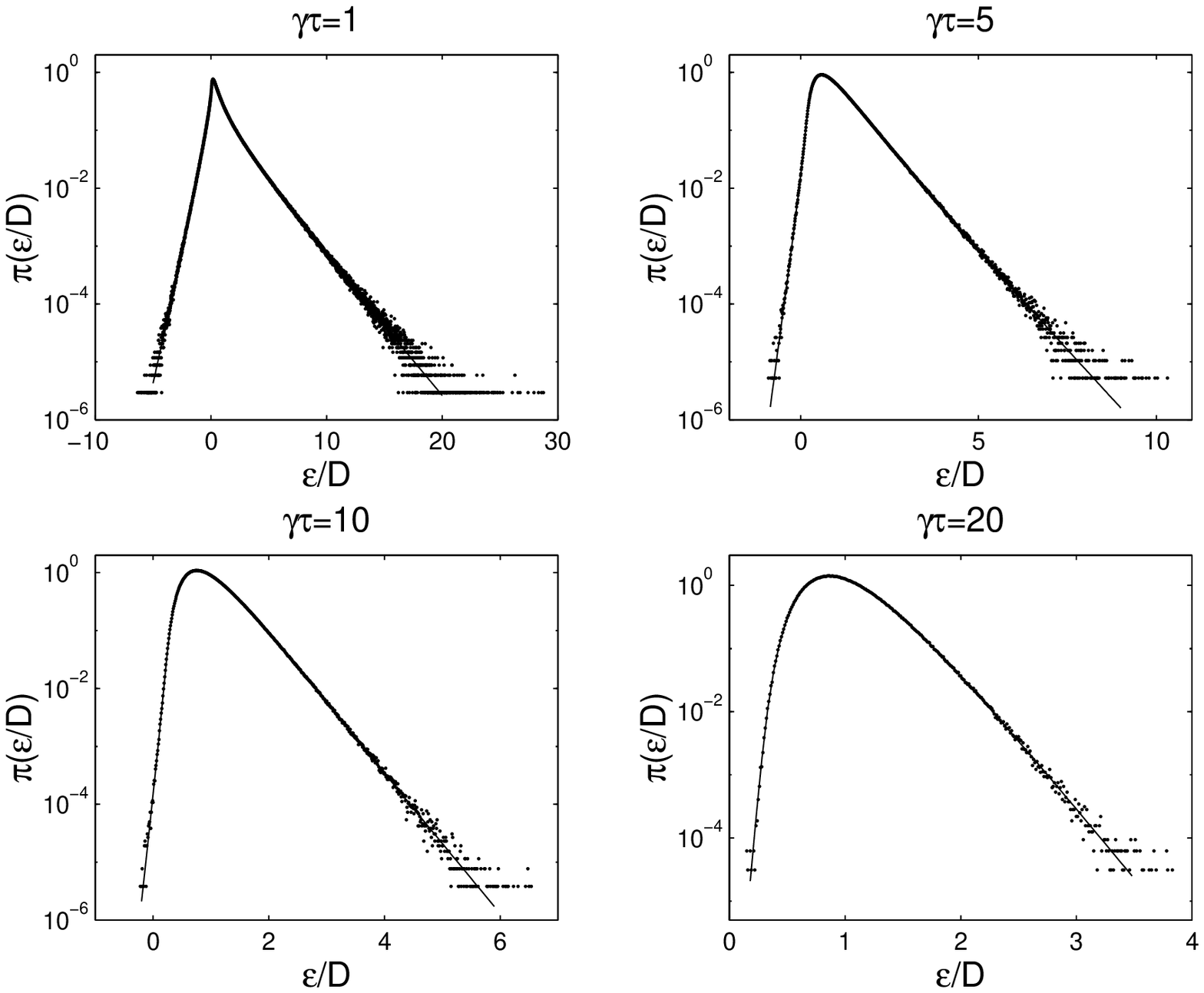}}
\end{center}
\caption{Semilog plots of the pdf of injected power $\pi(\teps)$ for
several values of $\tg=\ga\tau$. One plots together formula
\myref{fundamental} (line) and results of numerical simulation (dots).}\label{comparisons}
\end{figure}

It is easy to verify that neither $J$ nor $K$ have an exponential
leading term; thus, the large deviation function is easily computed as
\begin{align}\label{ff}
f(\teps)&=\left\{\begin{array}{ll}\displaystyle
-\frac{\ga}{4\teps}(\teps-1)^2 & \text{if $\teps\geq 1/3$}\\
\ga(2\teps-1) & \text{if $\teps\leq 1/3$}\end{array}\right.
\end{align}
This bipartite shape of the large deviation function is closely
related to the location of the saddle of the integration path, which
remains fixed in the complex plane as soon as $\teps<1/3$ (see figure
\ref{chemins}). These results are successfully compared with numerical
simulations, see fig. \ref{comparisons}.

\subsection{Discussion}

These results address naturally some questions, in particular
concerning the surprising structure of $f(\teps)$: this function
displays a second order singularity located at a odd $\teps=1/3$
value.

Let us first give some remarks on the shape of $\pi(\teps)$ (we
restrict the discussion henceforth to situations where
$\tau\gg\ga^{-1}$). This pdf is a
rather asymmetric curve, what can appear at first sight surprising : as the renewal of the noise $\psi$ is independent of the particle velocity $v$, occurrences of positive or negative instantaneous power injection $\psi v$ are completely equiprobable. 
Actually,  the long time interval during which the mean is performed
is of crucial importance and is responsible for the peculiar shape of
the pdf; to understand this, let us consider an occurrence of a (rare)
large positive fluctuation of injected power: this occurrence
understands  that a favourable sampling of the noise is realized, so
that very often the noise gives energy  to the particle. Consequently,
during the process, the energy has a global tendency to increase, as
well as typical values of the velocity (despite the always acting dissipation); as the injected power is
directly proportional to the velocity, one sees that the direct effect
of the  positive injection of energy is to enhance typical values of
$v$ implied in the evaluation of $\eps$: this favourable feedback makes finally the occurrence of the considered fluctuation more likely, since less efficiency of the noise is globally required to generate the fluctuation.

On the contrary, let us consider a (large) negative fluctuation: the
scenario is here inverted, since the typical velocity will certainly
decrease during the mean process : to ensure the large expected value
of power ceded back to the bath, one is then compelled to begin the
motion with a large kinetic energy, what is exponentially unprobable :
the feedback is here clearly unfavourable, and diminishes
comparatively the frequency of such occurrences.

This primordial role of the initial energy on the negative power
injection mechanism gives actually also the explanation for the
presence of the negative tail in the large deviation function of
$\pi(\teps)$: each fixed initial velocity is surely relaxed within a
characteristic time $\propto \ga^{-1}$,  but this time is longer the
higher initial velocity $v_0$; thus, when this velocity is
statistically distributed according a Gaussian, rare large initial
velocities construct the negative tail of the distribution as well as the
associated large deviation function, which precisely characterizes
rare events. It is worth to note that the specific negative tail due to
the thermalization of $v_0$ begins actually at the value $1/3$, and
 thus affects substantially also the positive part of the
distribution.

This singularity of the large deviation function can be interpreted as
a phase transition, if one writes $\ft(\la)$ as a
functional integral over the realizations of the noise
$\psi$. Integration on $v_0$ gives
\begin{align}
\ft(\la)&\propto\int\mathcal{D}\psi\exp\left(-\frac{1}{4}\int_0^\tau\psi^2-\frac{\la
D}{2}\int_0^\tau
dt\ dt'\
\psi(t)\psi(t')e^{-\ga|t-t'|}+\frac{(\la D)^2}{2\ga}\left[\int_0^\tau dt\
\psi(t)e^{-\ga t}\right]^2\right)
\end{align}
and this expression can be viewed as  a configurational partition
function of an unidimensional line $\psi(t)$ of length $\tau$
confined in a quadratic potential $\mathcal{V}[\psi]\propto \int \psi^2$, with
short range homogenous interactions (second term) and an
additional local destabilization term (third term), which is a direct consequence of the thermalization of
$v_0$. If $\la$ is positive and too large, the combined effects of the
 short range interaction term and the third term --which can
be viewed more or less as an inverted parabolic potential acting in
the vicinity of 
the $t=0$ end of the chain only-- destabilizes the chain which is no
longer confined by $\mathcal{V}$.

We would like to make here a little mathematical digression. One can compute the characteristic function from the preceding formula. This leads to the following formal expression
\begin{multline}
\ft(\la)=\exp\left(-\demi\sum_{n=1}^{\infty}\log\left(1+\frac{2\tla}{1+x_n^2}\right)\right.\\
\left. -\demi\log\left(1-4\tla^2\sum_{n=1}^{\infty}\frac{x_n^2}{1+x_n^2}\frac{1}{2+\tg(1+x_n^2)}\frac{1}{1+2\tla+x_n^2}\right)\right)
\end{multline}
where $x_n$ is solution of the implicit equation $\tg x_n+2\text{Atan} x_n=n\pi$.  That this involved expression is exactly the simple formula \myref{pila} is quite amazing, and one can derive some nontrivial mathematical relations from this connection; to mention but a few, one has for instance
\begin{align}
\sum_{n=1}^{\infty}\frac{1}{1+x_n^2}=\frac{\tg}{2}
\end{align}
a relation which is easily checked for $\tg\rightarrow 0$ ($x_1\sim\sqrt{2/\tg}$) and $\tg\rightarrow \infty$ (the sum becomes a Riemanian sum).

\bigskip

To conclude this paragraph, it is important to note that, in contradiction with
intuition, initial conditions can have deep consequences on the shape
of large deviations functions, even if the process displays finite
time correlations, provided that these initial conditions are
statistically distributed over an unbounded interval, what is almost
always the case if for instance stationary processes are considered.

\subsection{Dissipated Power}

\begin{figure}
\begin{center}
\resizebox{8cm}{!}{\includegraphics{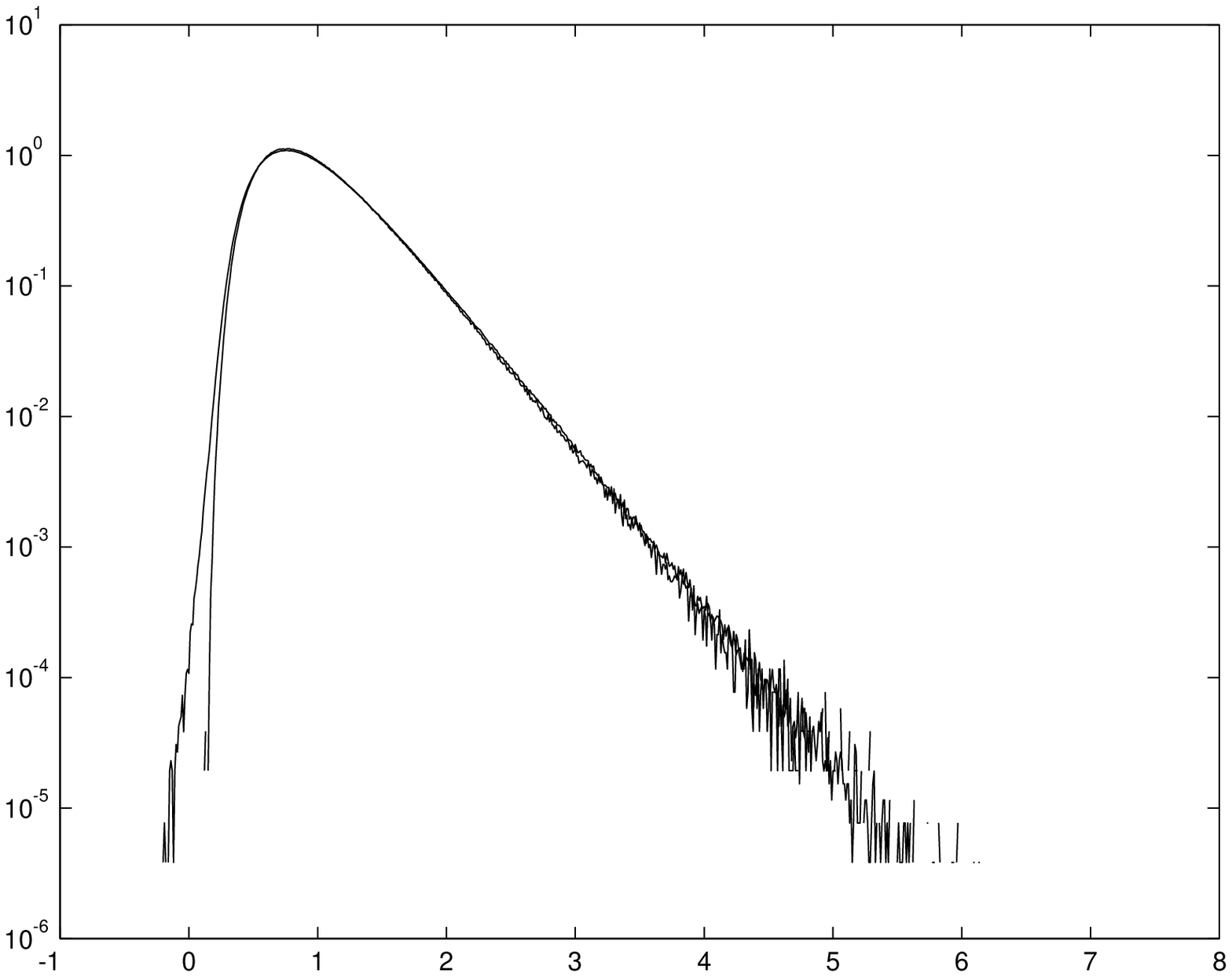}}
\end{center}
\caption{Injected and dissipated power pdf for the free Brownian
motion ($\ga\tau=10,D=1$). The two distributions are very similar but
in the vicinity of zero, partly due to the constraint $\ga v^2\geq0$,
whereas $\psi v\gtrless 0$.}
\label{dissinj}
\end{figure}

The pdf of dissipated power can also be computed along the same line
of reasoning. One gets for the characteristic function of the
stationary process:
\begin{align}
\left\langle \exp -\la\ga\int_0^\tau dt\ v^2(t)\right\rangle&=e^{\tg/2}\left(\cosh\eta\tg+\eta\sinh\eta\tg\right)^{-\demi}
\end{align}
and one sees that there is not any positive cut in that case. As a
result, the large deviation function of the \textit{stationary} dissipated power
$\de=\frac{1}{\tau}\int_0^\tau \ga v^2$ is easily derived as
$f_{dissip}(\de)=f_0(\de)$. Consequently, one expects the pdf of
dimensionless injected and dissipated power to be (in the large
$\tau$ limit) similar but in the zero injection (or dissipation)
region. This is effectively the case, as shown on figure \ref{dissinj}.

\subsection{Note on the Fluctuation Theorem}

As explained in the introduction, our model is a good system to test the possible universality of the
\textit{conclusions} of the (Evans-Cohen-Morris) fluctuation theorem, since the exact result is at hand. From \myref{ff}, one has
\begin{align}\label{ftendefaut}
\rho(\teps)\equiv\frac{1}{\tau}\log\frac{\pi(\teps)}{\pi(-\teps)}&\sur{\sim}{\tau\rightarrow\infty}\left\{\begin{array}{ll}
\displaystyle 4\ga\teps& \mbox{if $\teps<1/3$}\\
\displaystyle\frac{7}{4}\ga\teps+\frac{3}{2}\ga-\frac{\ga}{4\teps}& \mbox{if $\teps>1/3$}
\end{array}\right.
\end{align}
This function is clearly not a straight line, as it would be if the FT
held here (not even the slope at $\teps=0$ is in accordance with the FT, which would predict $\rho(\teps)=\ga\teps$). Thus we exhibit here an example where the conclusions of
the theorem are not verified (due to the simple fact that the situation
considered here does not fulfill the hypotheses required for the
Fluctuation Theorem to hold). On this ``negative'' result can we make
two comments: first, it is not contradictory with those of Kurchan
\cite{kurchan}. One looks for the power injected by a random fluctuation in
a dissipative system, whereas the
Fluctuation Theorem established by Kurchan considers the power
injected by an \textit{external} operator in a system in equilibrium
with a thermostat. We already noticed that the first situation is much
more appropriate to describe realistic systems driven far from
equilibrium. Second, formula \myref{ftendefaut} illustrates well the
fact that Fluctuation Theorem seems to hold in so large a number of
experimental situations, as explained in \cite{aumaitre}: in the
vicinity of $\teps=0$, $\rho(\teps)$ must always have  a straight line
behaviour, as a consequence of the large deviation law; on the other
hand, as large negative values of $\teps$ are extremely unprobable
when $\tau$ is large, it becomes practically unpossible even to only
measure $\rho(\teps)$ for large $\teps$ and large $\tau$ with enough
statistical resolution: possible deviations from the straight line are
just even not measurable. In our case, crossover occurs for $\teps=1/3$
and for this value, $\pi\propto\exp(-5\ga\tau/3)$ which is of order
$10^{-8}$ only if $\ga\tau=10$\ldots Our model is thus a good
illustration in favour of arguments given in \cite{aumaitre} against
an universal applicability of conclusions of the Fluctuation Theorem.

\begin{figure}[h]
  \begin{center}
    \resizebox{6cm}{!}{\includegraphics{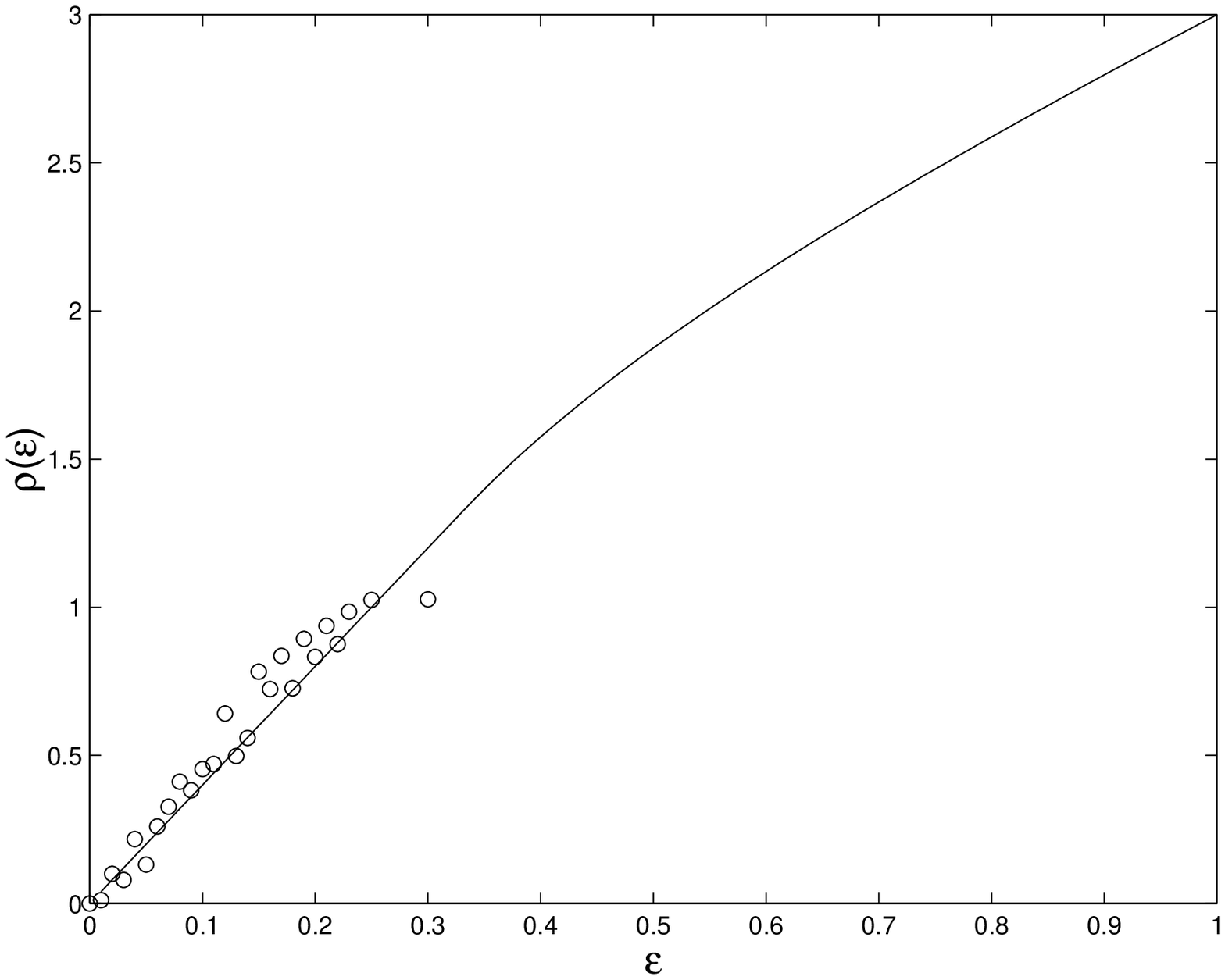}}
    \caption{Function $\protect\tau^{-1}\protect\log[\protect{\pi}(\protect\teps)/\pi(-\protect\teps)]$ for
$\ga\tau=8$. Circles come from numerics with two
millions points of statistics, and stop before the crossover of the
theoretical curve (solid) due to the lack of negative points with
large absolute value.}
    \label{pisurpi}
  \end{center}
\end{figure}

\section{Confined Brownian Motion}

The reasoning concerning the asymmetry of $\pi(\teps)$ suggests that certain characteristics of this pdf seem ---in the limit of large $\tau$ only--- to be independent of the details of the particle dynamics, since it is based only on considerations on energy and its conservative character. These observations led us to infer a possible insensivity of $\pi$ with respect to other microscopic times than $\ga^{-1}$ ($\ga^{-1}$ itself cannot be neglected, since this time plays a role in the process of dissipation of energy; and indeed, the curves for different $\ga$ have different and non superposable shapes), in cases where the initial system would have been complexified. Thus, we considered several confined Langevin systems:
\begin{align}
  \ddot{x}+\ga \dot{x}+V'(x)&=\psi(t)\label{clangevin}\\
<\psi(t)\psi(t')>&=2D\de(t-t')
\end{align}
where we used for $V(x)$ an harmonic potential $V(x)=\demi\om^2x^2$, a
non linear ``hard'' potential $V(x)=\demi\om^2x^2+\frac{\al}{4}x^4$
($\al>0$), a non linear asymmetric ``soft'' potential
$V(x)=\om^2(\al^2e^{-x/\al}+\al x)$, and also a bistable $\varphi^4$
potential $V(x)=V_b(x^2-1)^2$. We numerically calculated for all these
potentials (as well as again the free case for comparison) the pdf
$\pi(\teps)$, for large  values of $\tau$ ($\ga=1$ for
convenience).
\begin{figure}[h]
  \begin{center}
    \resizebox{10cm}{!}{\includegraphics{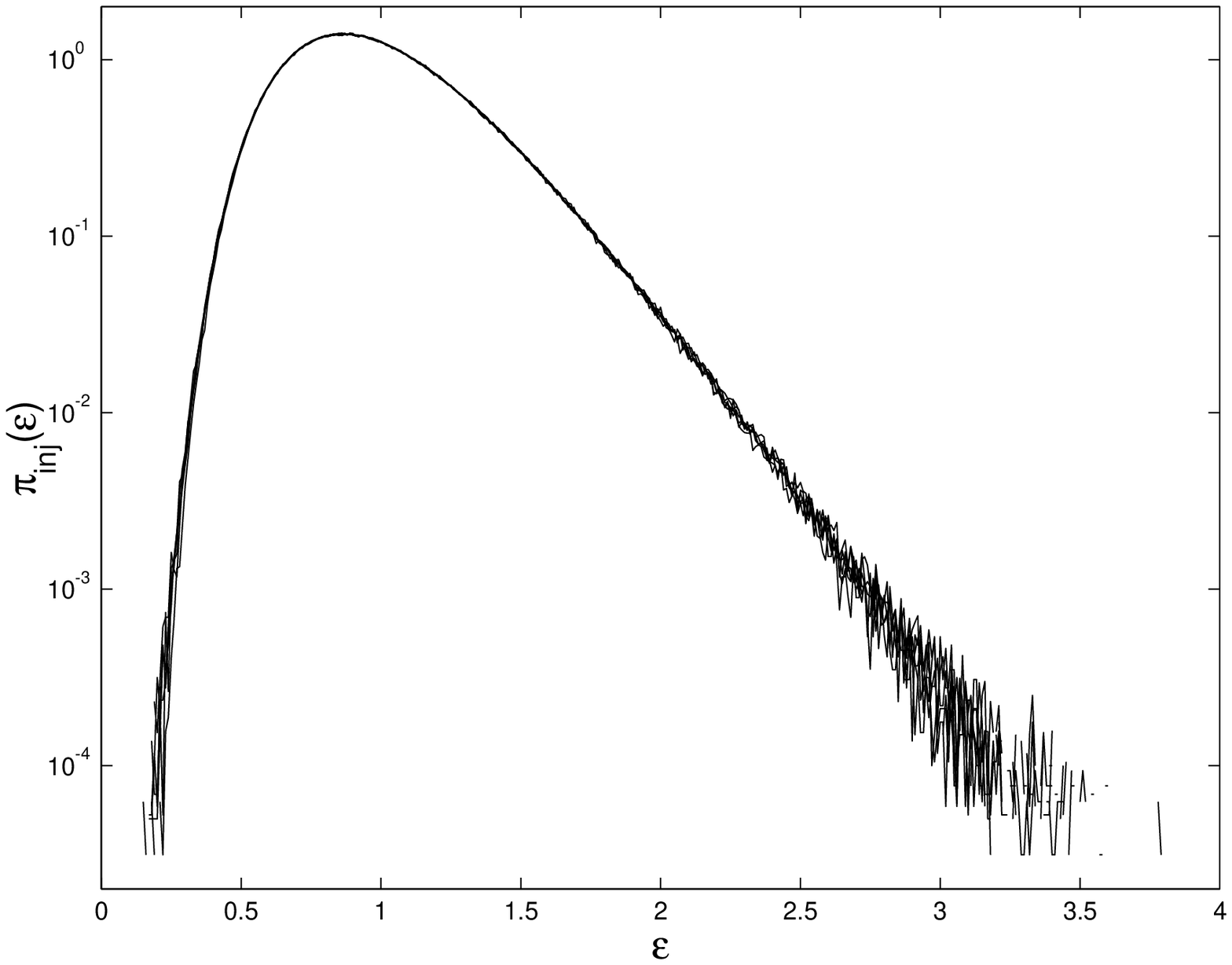}}
    \caption{Injected power pdf for several trapped Langevin particles with
$D=1$, $\tau=20$ and  $\ga=1$ : three harmonic potentials (free
motion $\om=0$, $\om=1$ and
$\om=10$), hard and soft nonlinear potentials ($\om=1$,$\al=3$ in both
cases), bistable $\varphi^4$-potential with $V_b=1$ (strong
anharmonicity is explored in this case -- see details in text).}
    \label{constatations}
  \end{center}
\end{figure}

 The results (see fig. \ref{constatations}) are
extraordinarily surprising, since  one cannot
distinguish the different curves from each other!  One must keep in
mind the fact that the dynamical behaviour of $x(t)$ is completely
different in all these cases: fully isochronic or anisochronic
oscillatory, overdamped, bistable, diffusive, these different Brownian
motions lead all to apparently the \textit{same} curve, a coincidence
which goes beyond all expectations.

To understand this phenomenon, let us look at the characteristic
function  $\ft_{\rz}(\la)$, where $\rz=(x_0,v_0)$ designates the
initial conditions. Of course, one cannot compute it exactly, since in
general the dynamics is nonlinear. Nevertheless, it is possible to
express it in a fruitful form.

From \cite{zinnjustin}, one can derive the path integral
representation of the solution of the Kramers equation:
\begin{align}
P_\ga(\ru,\tau|\rz,0)&=e^{\tg/2}\times\int_{\mathbf{r}(0)=\rz}^{\mathbf{r}(\tau)=\ru}\hspace{-0.5cm}[\mathcal{D}x]\exp\left(-\frac{1}{4D}\int_0^\tau
dt\ [\ddot{x}+\ga\dot{x}+V'(x)]^2\right)
\end{align}
(the index $\ga$ recalls the value of the damping).
As before, one gets from this
\begin{align}
\ft_{\rz}(\la)=\int d\ru \int_{\mathbf{r}(0)=\rz}^{\mathbf{r}(\tau)=\ru}\hspace{-0.5cm}[\mathcal{D}x]\exp\left(\frac{\tg}{2}-\frac{1}{4D}\int_0^\tau
dt\ [\ddot{x}+\ga\dot{x}+V'(x)]^2-\la\int_0^\tau \dot{x}[\ddot{x}+\ga\dot{x}+V'(x)]\right)
\end{align}
After some simple manipulations, one can recast this into
\begin{align}
\ft_{\rz}(\la)=e^{\frac{\tg}{2}(1-\eta)}\int d\ru P_{\ga\eta}(\ru,\tau|\rz,0)\times \exp\left(-\frac{\ga}{2D}(\tla+1-\sqrt{1+2\tla})(E_1-E_0)\right)
\end{align}
where $E_i=\demi \dot{x}_i^2+V(x_i)$ (we recall $\tla=2D\la/\ga,
\eta=\sqrt{1+2\tla}$). This formula is useful, since the propagator $P_{\eta\ga}$
goes exponentially fast to the equilibrium value
$P_{eq}(\ru)=\sqrt{\ga\eta/(2\pi D)}U_{\eta}^{-1}\exp(-\ga\eta E_1/D)$ where $U_{\al}\equiv\int
dx_0 \exp(-\ga\al V(x_0)/D)$ is
the configurational partition function.
Consequently, for each fixed value of $\rz$, one has
the true  equivalence
\begin{align}
\ft_{\rz}(\la)&\sim \sqrt{\frac{\ga\eta}{2\pi D}}U_{\eta}^{-1} e^{\frac{\tg}{2}(1-\eta)}\int d\ru
\exp\left(-\frac{\ga}{2D}[(\tla+1+\eta)E_1-(\tla+1-\eta)E_0]\right)
\end{align}
and one sees that there is no problem of convergence in the integral
for any real value of $\tla$ such that $\tla>-1/2$ (for $\tla<-1/2$
there is always a cut due to the presence of a square root in $\eta$).
Thus, one extracts exactly the leading exponential term from the
preceding formula as
\begin{align}
\log\ft_{\rz}(\la)&\sim  \frac{\tg}{2}(1-\eta)
\end{align}
and shows in the same time that the large deviation function of
$\pi_\rz(\teps)$ is
\begin{align}
f_\rz(\teps)=-\frac{\ga}{4\teps}(\teps-1)^2\times\theta(\teps)
\end{align}
\textit{irrespective of the precise form of the potential}.

Let us now look at  $\pi(\teps)$: its characteristic function can be
written 
\begin{align}
\ft(\la)=\sqrt{\frac{\ga}{2\pi D}}\frac{e^{\frac{\tg}{2}(1-\eta)}}{U_1}\int d\rz d\ru
P_{\ga\eta}(\ru,\tau|\rz,0)\times
\exp\left(-\frac{\ga}{2D}[(\tla+1-\eta)E_1-(\tla-1-\eta)E_0]\right)
\end{align}
A priori, one must take care of the fact that the equivalence
$P(\ru,\tau|\rz,0)\sim P_{eq}(\ru)$ is not reached uniformly with
respect to $\rz$, as already noticed. But, if one inspects the free
Brownian case, for which the exact result is computed, this
replacement is finally equivalent to neglect exponentially small
corrections (terms like $\tanh\eta\tg$ replaced by $1$ for instance);
this is precisely this approximation which leads to formul\ae\ 
(\ref{j2},\ref{k2}) from (\ref{j1},\ref{k1}): the only limitation is
that the resulting formul\ae\  are not uniformly valid in the $\teps$
space. But the associated large deviation function is unaffected by
these corrections.

We can assume that this scenario is still correct in the general
case, an assumption which is very reasonable indeed. Thus, up to
exponentially vanishing factors, one has
\begin{align}
\ft(\la)&\sim \frac{\ga\sqrt{\eta}}{2\pi D}\times\frac{e^{\frac{\tg}{2}(1-\eta)}}{U_1 U_{\eta}}\int d\rz d\ru
\exp\left(-\frac{\ga}{2D}[(\tla+1+\eta)E_1-(\tla-1-\eta)E_0]\right)\\
&=\sqrt{\frac{4\eta}{(\eta+1)^2-\tla^2}}\times e^{\frac{\tg}{2}(1-\eta)}\times\frac{U_{(\tla+1+\eta)/2}U_{(\eta+1-\tla)/2}}{U_1
U_{\eta}}
\end{align}
and it is easily seen that, again, the $\rz$ integral
diverges when $\tla$ approaches $\tla_+=4$, pointing out  the probable
beginning of a real cut, already encountered when $V=0$. Noticing that
the leading exponential term is $\exp(\frac{\tg}{2}(1-\eta))$,
also insensitive to the presence of a pinning potential, one deduces
that \textit{also in this case the large deviation function is given
by \myref{ff}}. Of course, fully mathematical precision is not given
here, but we think that convincing arguments are nevertheless given in
favour of our result.

Concerning the prefactor of the large deviation function, it is
interesting to mention that it keeps in general a dependence on the
form of the pinning potential, through the functions $U$. 
 In fact, formulas like (\ref{j2},\ref{k2}) can
themselves be simplified; for instance
\begin{align}
J(\teps)\sim \left(\frac{4\tg}{\pi(3\teps-1)(\teps+1)^3}\right)^{\demi}
\end{align}
(we did not propose this equivalence previously, for it is not correct
in the vicinity of $\teps=1/3$, unlike formula \myref{j2}\ldots). For
the general case with a potential $V$, this ``supersimplification'' gives
\begin{align}
J(\teps)\sim\left(\frac{4\tg}{\pi(3\teps-1)(\teps+1)^3}\right)^{\demi}\times\frac{U_{(\teps^{-1}+1)^2/4}U_{(3-\teps^{-1})(1+\teps^{-1})/4}}{U_1U_{\teps^{-1}}}
\end{align}
Thus, the pdf associated with different potentials do not exactly coincide at large $\tau$, except at the value $\teps=1$. But, at the level of the large deviation function, the universality of $f(\teps)$ is reached. The combination of these two points explains probably the remarkable coincidence of the different curves on the figure \ref{constatations}.

\section{Conclusion}

In this paper, we considered the Langevin equation as a dynamical evolution of a simple dissipative system driven by an external forcing, and computed the probability density function of the time-averaged injected power in the permanent and non permanent regimes. We showed that the associated large deviation functions are different, in particular a negative tail exists only in the permanent regime. We explained the origins of this discrepancy and highlighted the role of the rare but very energetic initial conditions; we showed also that the system considered here does not verify the so-called Fluctuation Relation $f(\teps)-f(-\teps)=\ga\teps$, even in the vicinity of $\teps=0$, indicating thence that any attempt to enlarge careless the applicability of the conclusions of the Fluctuation Theorem to dissipative systems is vain.

We considered thereafter Langevin equations with pinning potential, and showed that the associated large deviation functions are completely insensitive to the potential (but not the pdf itself): this result appears to be a good indication that large deviation functions could be an appropriate tool to characterize well general properties of systems beyond some peculiar irrelevant details which faded away through the process of averaging. In our case, the function $f$ tells us something global associated with the energy transfer throughout the Brownian particle, irrespective to the precise dynamics of each geometrical configuration.

A natural extension will be to consider stochastic coloured dynamics, i.e. noises with time correlation. We want to test the ``solidity'' of $f$ with respect to correlations arising in the external forcing. Moreover, we can address the question of the generality of the observed non-analyticity of the large deviation function in systems in permanent regime where the velocity field is unbounded: from our analysis of the Langevin system, we think that such anomalies could be widely encountered in dissipative far-from-equilibrium systems. An interesting perspective is besides to consider extremely correlated noises, in order to check in this context the ideas exposed in \cite{bhp,bhp2}, where some conjectures are made on the limiting behaviours of the pdf of global variables in highly correlated systems.

\section{Acknowledgments}

I am much indebted to S.Fauve, R.Labb\'e, S.Auma\^\i tre, F.P\'etrelis, R.Berthet,
N.Mujica, R.Wu\-nen\-bur\-ger and B.Derrida for fruitful discussions. I acknowledge also a referee for having suggested me an alternative and fruitful technical approach of the computation.

\end{document}